\documentclass[twocolumn,  superscriptaddress,  amsfonts,floatfix]{revtex4} 

\usepackage{color}

\usepackage{graphicx}
\usepackage{dcolumn}
\usepackage{bm}

\begin{document}

\preprint{APS/123-QED}

\title{ 
Competition between Geometrical Frustration and the Kondo Effect in CePdAl Revealed by High-Resolution Magnetization
 \color{black}
}

\author{Yusei Shimizu }
\email{yshimizu@issp.u-tokyo.ac.jp }
\affiliation{Institute for Solid State Physics, The University of Tokyo, Kashiwa, 277-8580, Japan}
\author{Shota Nakamura}
\affiliation{Nagoya Institute of Technology, Aichi, Nagoya 466-8555, Japan}
\author{Yoichi Ikeda} 
\affiliation{Institute for Materials Research, Tohoku University, Sendai, 980-8577, Japan}
\author{Yohei Kono}
\affiliation{  School of Science and Engineering,  
 Chuo University, Kasuga, Bunkyo-ku, Tokyo, 112-8551, Japan}
\author{Shunichiro Kittaka} 
\affiliation{Department of Basic Science, The University of Tokyo, Meguro, Tokyo 153-8902, Japan}
\author{Toshiro Sakakibara}
\affiliation{Institute for Solid State Physics, The University of Tokyo, Kashiwa, 277-8580, Japan}
\author{Yosikazu Isikawa}
\affiliation{Graduate School of Science and Engineering, Univ. of Toyama, Toyama, Toyama 930-8555, Japan}

\date{\today}

\begin{abstract}
CePdAl is a heavy-fermion compound with a quasi-kagome structure, where geometrical frustration competes with the Kondo effect. Using a high-sensitivity magnetometer, we observe clear first-order metamagnetic transitions without magnetization plateaus in CePdAl at 80 mK, indicating a lifting of frustration through the suppression of the Kondo effect and spin flips of ordered moments under fields. An anomaly in the nonlinear magnetic susceptibility at 4.3 T suggests nondipolar correlations in the polarized paramagnetic state.
 Furthermore, we found no evidence of non-Fermi-liquid behavior in the high-field region, where the antiferromagnetic order is completely suppressed. These findings establish the essential  thermodynamic constraints for understanding the field-induced spin-liquid state in CePdAl.
  \color{black}
\end{abstract}

\maketitle


The exploration of unconventional quantum electronic states induced by geometrical  frustration has been  a hot topic in condensed matter physics. Geometrical frustration suppresses a magnetic ordering  and can lead to a variety of novel quantum phases, including spin liquids \cite{Lee_2008, Balents, FrustratedMagnetism}, spin ice \cite{Ramirez_1999, Ramirez_frustratedMag, Bramwell}, dimer orders \cite{FrustratedMagnetism}, and partially ordered states \cite{Mentink, Lacroix, Keller, Donni}. Despite extensive studies, however, the nature of the ground state in geometrically frustrated quantum many-body systems remains  nontrivial. The Kondo effect is a prototypical quantum many-body phenomenon in which a localized $f$-electron spin is screened by conduction electrons to form a 
 Kondo spin-singlet state. Strongly correlated electron systems in which the Kondo  effect  competes with geometrical  frustration provide an attractive basis for uncovering hidden quantum states  that are distinct from those in  localized spin systems.
 \color{black}

CePdAl is an unconventional heavy-fermion compound in which the Kondo effect ($T_{\rm K}$ = 5 K) \cite{Kitazawa, Isikawa_JPSJ_1996} competes with geometrical frustration to give rise to a rich variety of physical properties. This material crystallizes in a noncentrosymmetric hexagonal structure (space group $\#189$, $P\bar{6}2m$) \cite{Xue, Hulliger} and exhibits an antiferromagnetic (AF) order below $T_{\rm N}=2.7$ K \cite{Kitazawa, Isikawa_JPSJ_1996}.
 In this structure, three crystallographically equivalent Ce sites form  a quasi-kagome structure composed of triangular arrangements of Ce ions, which lead naturally to geometrical  frustration.  Magnetic moments at the nearest-neighbor Ce sites are coupled ferromagnetically ($J_{1}$), whereas the moments at  next-nearest-neighbor Ce sites interact antiferromagnetically ($J_{2}$),  
  within a quasi-two-dimensional kagome network of Ce ions stacked along the $[0001]$  direction   \cite{Donni}.
   In the paramagnetic (PM) state, CePdAl exhibits strong magnetic anisotropy with the easy-magnetization axis along the $[0001]$ direction \cite{Isikawa_JPSJ_1996}. Below $T_{\rm N}$, neutron-scattering experiments have revealed an incommensurate magnetic structure characterized by the propagation vector $\bm{q}=(\frac{1}{2},0,\tau)$ with $\tau\sim0.35$ \cite{Keller, Donni, Prokes}, where the magnetic moments at the Ce(1) and Ce(3) sites are sinusoidally modulated.
 Remarkably,   magnetic moments   at the Ce(2) sites  remain partially disordered  below $T_{\rm N}$ \cite{Keller, Donni, Prokes}. This partial disorder is believed to relieve geometrical frustration through Kondo screening \cite{Keller, Donni},   and this  scenario is also   supported by $^{27}$Al NMR measurements down to 30 mK \cite{Nishiyama, Oyamada}.
 Under  fields applied along  $B\parallel[0001]$, CePdAl undergoes three successive metamagnetic transitions \cite{Goto}, resulting in a  rich  magnetic $B$–$T$ phase diagram \cite{Zhao, Mochi_JPSJ_2017, Lucas_PRL_2017, Zhang_PRB_2018}. This behavior has been attributed to the  geometrical frustration as the Kondo effect is progressively suppressed by the applied magnetic field. Spin-liquid-like behavior has been proposed in the vicinity of the field-induced phases \cite{Lucas_PRL_2017, Ishant_PRB_2025, Zhang_PRB_2018}, as well as in the high-pressure regime near the AF quantum critical point ($P_{c}\sim0.8$ GPa) \cite{Prokes_2015, Zhao_Nature} and in the Ni-substituted  CePd$_{1-x}$Ni$_x$Al \cite{Isikawa, Fritsch, Majumder_PRB_2022,  Ishant_2024}.
However, a unified  physical picture of how such a spin-liquid state develops has remained elusive.

\color{black}

To gain insight into the magnetic ground-state properties of CePdAl, we performed high-resolution DC magnetization measurements at temperatures down to 70 mK. 
This study provides a new thermodynamic perspective by revealing  nondipolar correlations without long-range ordering. Furthermore, our measurements demonstrate the absence of  quantum critical behavior in the high-field region, 
 where a spin-liquid state has been proposed \cite{Zhang_PRB_2018}. These findings establish new experimental foundations  for understanding  the nature of the quantum ground state  at the boundary between the magnetic order and disorder in  CePdAl.
\color{black}

\begin{centering}
\begin{figure}[!htb]
\includegraphics[width=9.2cm]{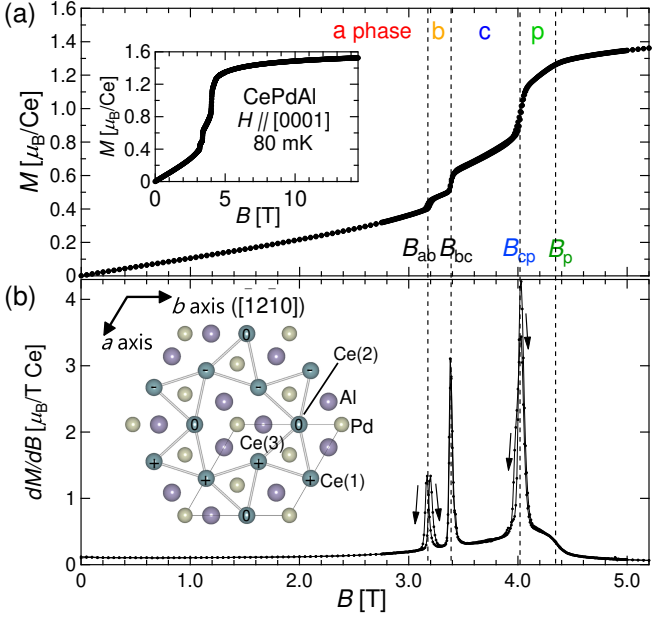}
\caption{ (a) The magnetization $M(B)$ of CePdAl for $B$ $||$ $[0001]$
 at 80 mK. The inset shows $M(B)$ in fields up to 14.5 T.
   (b) The differential magnetization $dM/dB$ of CePdAl for $B$ $||$ $[0001]$ at 80 mK, 
 where arrows indicate increasing and decreasing processes.\color{black}
The inset shows the crystal structure of CePdAl in the basal plane, with  ordered moments at  the  Ce(1) and Ce(3) sites
 and disordered moments at the Ce(2) sites, where  $+$, $-$, and $0$ indicate up, down, and disordered spins, respectively
  \cite{Donni}. 
\color{black}
}
\end{figure}
\end{centering}


Single crystals  were grown in an induction furnace using the Czochralski method \cite{Isikawa_JPSJ_1996}. We determined the crystal orientation by using the Laue x-ray method, and we cut out a 13.5 mg sample for the DC magnetization measurements using a spark cutter.  We performed high-resolution magnetization measurements using a home-built capacitive-detection Faraday magnetometer \cite{Sakakibara_JJAP_1994, Shimizu_RSI_2021} installed in a $^{3}$He–$^{4}$He dilution refrigerator, which allowed measurements down to 70 mK. For the magnetization measurements, we applied a uniform magnetic field along the easy-magnetization axis (the hexagonal $[0001]$ direction) up to 14.5 T, with a field gradient $dB_z/dz = 2$ T/m. We performed magnetization measurements above 2 K using a commercial SQUID magnetometer.


Figures 1(a) and 1(b) \color{black}
 show  the magnetization curve $M(B)$  and   differential magnetization, $dM(B)/dB$, respectively  measured at 80 mK
  along  $B$ $||$ $[0001]$.  The $M(B)$ curve exhibits clear magnetization jumps (step-like behavior) at $B_{\rm ab}=3.2$ T, $B_{\rm bc}=3.4$ T, and $B_{\rm cp}=4.0$ T; correspondingly, $dM(B)/dB$ shows sharp peaks at these transitions.
The observed step-like behavior clearly indicates the occurrence of first-order phase transitions at  $B_{\rm ab}$, $B_{\rm bc}$, and $B_{\rm cp}$. We observed no additional metamagnetic transitions above $B_{\rm cp}$, and the magnetization does not become fully saturated, remaining below the value calculated from crystalline electric-field (CEF) effects \cite{Mochi_JPSJ_2017}. This observation implies that Kondo screening is not fully destroyed even above 4 T, although the Kondo effect is progressively suppressed by the applied  field. Here, the CEF levels of the $4f^{1}$ configuration are split into three Kramers doublets \cite{Isikawa_JPSJ_1996, Mochi_JPSJ_2017, Woitschach_Neutron,   Romero_CEF}.

\begin{figure}[!htb]
\includegraphics[width=9.2cm]{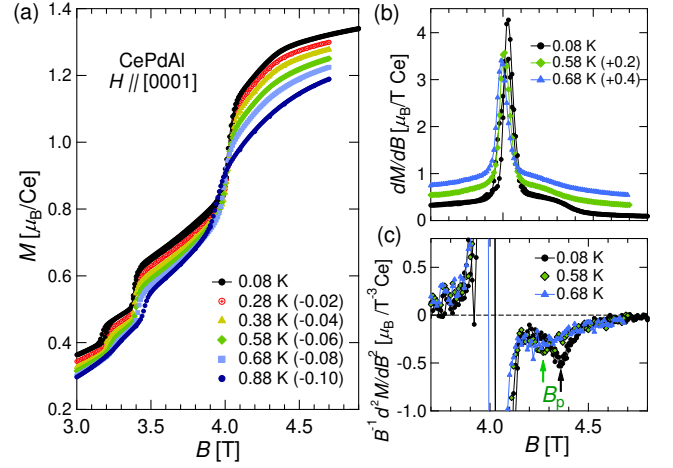}
\caption{(a) The magnetization curves $M(B)$ of CePdAl, measured at $T = $ 0.08, 0.28, 0.38, 0.58, 0.68, and 0.88 K.
  Here, the plotted values are shifted vertically for clarity, with the vertical shifts indicated in parentheses.
   (b) The differential magnetization, $dM/dB$, at 0.08, 0.58, and 0.68 K.   (c) The quantity $\chi_{3} \simeq B^{-1} d^2 M/dB^2$ for data obtained at 0.08, 0.58, and 0.68 K. 
    The upper arrows indicate the nonlinear susceptibility anomalies at $B_{\rm p}$. 
   \color{black}
}
\end{figure}

To understand the magnetic properties of CePdAl, it is essential to clarify the quantitative magnetization jumps at each first-order transition. Just below the lowest metamagnetic transition at $B_{\rm ab}=3.2$ T, the induced magnetization reaches approximately one-third of the induced  moment  at $B_{\rm cp}=$ 4.0 T  ($\sim1.2$ $ \mu_{\rm B}$/Ce).  \color{black} At $B_{\rm ab}$,  geometrical  frustration appears  \color{black} as the Kondo effect is suppressed by the applied  field. Notably, we observed magnetization jumps
 nearly  corresponding to $1/12$ and $2/12$ of the  saturated moment ($M_{\rm s}$)  at $B_{\rm ab}$ and $B_{\rm bc}$, respectively.
  Previous neutron-scattering experiments have reported no magnetic reflections other than $\bm{q}=(\frac{1}{2},0,0.35)$, which is observed at zero field. The intensity of $\bm{q}=(\frac{1}{2},0,0.35)$  reflection exhibits step-like decreases at both $B_{\rm ab}$ and $B_{\rm bc}$, but it persists in fields up to 4 T, whereas the  Bragg reflection of $(100)$ shows step-like increases \cite{Prokes}.
 The field-induced b and c phases can be explained  
 by the coexistence of AF order $\bm{q}=(\frac{1}{2},0,0.35)$ and another magnetic structure with  $\bm{q}= 0$ that  is stabilized in magnetic fields \cite{Prokes}.
 In this case,  several   Ce(2) sites  may become ordered  to form  the   $\bm{q}= 0$ order, while 
   other Ce(2) sites remain disordered,  playing  the role of a magnetic antiphase boundary between the above  magnetic structures. 
   As discussed later, the $M(T)$ curves differ from those of a simple ferromagnetic (FM)  order in the b and c phases,
   implying  ferrimagnetic-like  $\bm{q}= 0$ orders.  \color{black}
    Furthermore, assuming that the Ising moments at each Ce site cannot cant, the identical slope of 
 $M(B)$  [Fig. 1(a)] in the a, b, and c phases suggests that there exist partially disordered moments   in each phase.

\begin{centering}
\begin{figure}[!htb]
\includegraphics[width=8.9cm]{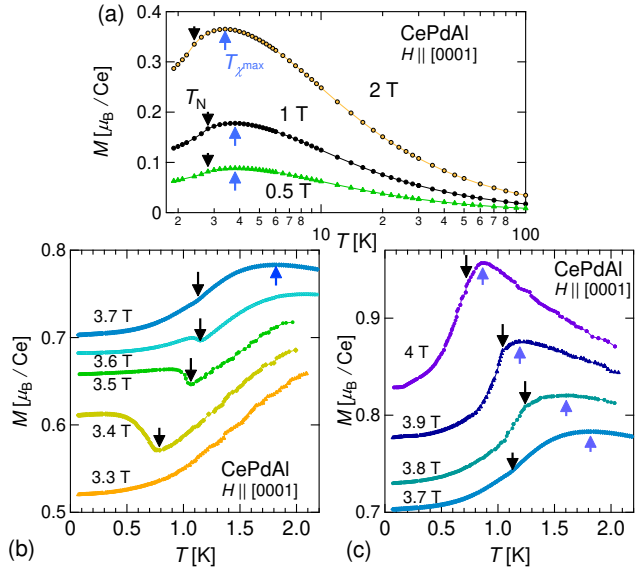}
\caption{(a) Temperature dependence of the magnetization $M(T)$ measured at 0.5, 1, and 2 T along $H$ $||$ $[0001]$ between 2 and 100 K on a logarithmic temperature scale.
  (b) Low-$T$ magnetization $M(T)$ below 2 K, measured along $H$ $||$ $[0001]$ at 3.3, 3.4, 3.5, 3.6, and 3.7 T, and (c) at 3.8, 3.9, and 4 T (with 3.7 T for comparison). The AF transitions and the susceptibility maxima are indicated by the upper and lower arrows, respectively. 
}
\end{figure}
\end{centering}

At $B_{\rm cp}\simeq 4.0$ T, the $M(B)$ curve exhibits a large magnetization jump, indicating a first-order phase transition at $B_{\rm cp}$. 
 When the magnetic moments in the in-plane down-spin chains undergo a spin-flip transition at $B_{\rm cp}$, a magnetization jump of $\frac{2}{3} M_{\rm s}$ is expected. The observed jump  is nearly half of this estimate, which can be attributed to the reduction of the net moments caused by the sinusoidal modulation along the 
 $\tau \sim $0.35  at the Ce(1) and Ce(3) sites.
  Importantly, the magnetization curve does not become fully saturated even above $B_{\rm cp}$ up to 4.3 T. This behavior is consistent with previous neutron-scattering results, which show that the  Bragg reflection  of  $(100)$ does not saturate in a similar field range (4.1–4.3 T) \cite{Prokes}. These facts indicate that Kondo screening persists even above $B_{\rm cp}$, as the slope of the  $M(B)$  curve remains comparable to those observed in the a and b phases.
 We  observed an anomalous change in the slope of the magnetization curve $M(B)$ (a kink-like behavior) around 4.3 T, where the differential magnetization $dM/dB$ [Fig. 2(b)] exhibits a shoulder anomaly
 \cite{EndNote_inhomogeneity}. 
Previous studies of the AC susceptibility, thermoelectric power, and Hall resistivity have also revealed an anomaly 
 at  4.3 T  \cite{Zhang_PRB_2018}, supporting its intrinsic nature.

We next focus on the temperature evolution of the $M(B)$ curves. Figures 2(a) and 2(b) show the $M(B)$ curves and the differential magnetization $dM/dB$, respectively, measured at several temperatures between 80 mK and 0.88 K. With increasing $T$, the peak in $dM/dB$ is gradually suppressed, while the shoulder anomaly becomes broader and eventually disappears around 0.7 K.
Figure 2(c) shows the quantity $   \chi_{3} \simeq \frac{1}{B}\frac{d^{2}M}{dB^{2}} $, where $M = \chi_{1} B + \chi_{3} B^3 +...$. We observe a dip anomaly 
  at $B_{\rm p}$ \color{black}
  with a negative  $\chi_{3}$ around the shoulder anomaly in $dM/dB$  [Fig.  2(c)]. 
 For the  metamagnetic transitions   at $B_{\rm ab}$, $B_{\rm bc}$ and $B_{\rm cp}$, \color{black}
 $\chi_{3}$  exhibits a sign change from positive to negative at increasing $B$. In contrast, only a negative  $\chi_{3}$ is observed for the shoulder anomaly at 4.3 T, indicating that the origin of this nonlinear susceptibility anomaly is qualitatively distinct from those associated with the metamagnetic transitions.

\begin{figure}[!htb]
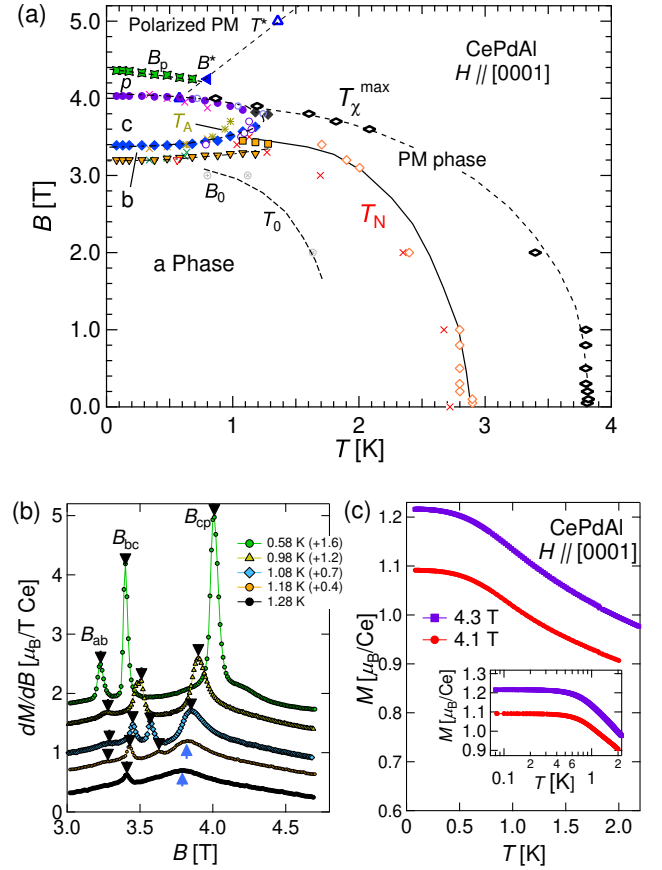

\includegraphics[width=8.9cm]{Fig4a_H-T_Phase_diagram_2ndProof.eps}
\includegraphics[width=8.9cm]{Fig4b_c_Layout2_RR.eps}
\caption{ (a) $B$-$T$ phase diagram of CePdAl   in fields applied along the hexagonal $[0001]$ axis, obtained from $M(B)$ (closed markers) and $M(T)$  (open markers).
The heat-capacity anomalies of crossovers $B^{*} $, $T^{*}$, $T_{\rm A}$, $T_0$,  $B_0$,  and phase transitions (cross markers) are   cited  from the Ref.   \cite{Mochi_JPSJ_2017}. The solid and dashed lines are guides to the eyes.
 (b) The differential magnetization $dM/dB$ at 0.58, 0.98, 1.08, 1.18, and 1.28 K with shifted values.
  The up and down arrows indicate transitions and crossover (susceptibility anomaly), respectively. 
(c) The magnetization $M(T)$, measured at 4.1 and 4.3 T. 
 The inset shows the corresponding semi-logarithmic plot.
 \color{black}
}
\end{figure}

 The newly observed shoulder-like anomaly (magnetization kink) in the $M(B)$ curve and 
   the accompanying  nonlinear susceptibility anomaly at $B_{\rm p}$ cannot be understood solely in terms of conventional  dipolar correlations. 
   In general, a kink anomaly in the magnetization of an ordered magnetic system is naturally associated with the canting or reorientation of magnetic dipole moments induced by an applied magnetic field. However, this scenario is unlikely for CePdAl. Because of its strong Ising anisotropy, the magnetic moments are already nearly polarized in   fields above $B_{\rm cp}$, making  canting or reorientation of the dipole moments unlikely.  Our high-resolution magnetization measurements demonstrate that the kink anomaly occurs without any thermodynamic phase transition at $B_{\rm p}$, as evidenced by $M(T)$ measurements discussed below. 
  The absence of symmetry breaking  across  $B_{\rm p}$ provides a crucial thermodynamic constraint,  indicating  that  the observed anomaly cannot be understood within a simple  dipolar picture alone.  The  third-order  nonlinear  susceptibility is a sensitive probe of  quadrupolar  correlations, as established by Morin $et$ $al$. \cite{Morin_PRB_1981}. The observed  nonlinear susceptibility anomaly suggests that  nondipolar correlations   develop  in the field-induced disordered state ($B_{\rm cp} < B  < B_{\rm p}$) in CePdAl.

 \color{black}

 Figure 3(a) shows the temperature dependence of the magnetization $M(T)$ in the high-$T$ region. Notably, the magnetic susceptibility exhibits a maximum at $T^{\chi}_{\rm max}$, reflecting  the characteristic temperature  of   frustration in CePdAl.   Figures 3(b) and 3(c) display the low-temperature $M(T)$ curves for CePdAl measured down to 70 mK under fields between 3.3 and 4 T. 
 We observed kink anomalies at the magnetic phase transitions 
  (down arrows). \color{black} 
  At 3.5–3.6 T, the $M(T)$ increases upon cooling below the magnetic transitions with clear kink anomalies, suggesting the increase of   magnetic moments in the c phase.   
  FM moments usually develop smoothly under magnetic fields, thus the c phase 
   is not  explained by the conventional FM $\bm{q} = 0$ state, implying an unconventional magnetic structure  with $\bm{q} = 0$ that stabilizes and coexists with $\bm{q}=(\frac{1}{2},0,\tau)$  in applied field. 
     At 3.7 T, the $M(T)$ shows a small kink at the magnetic transition,  whereas at 3.8 and 3.9 T,  $M(T)$ exhibits a pronounced  decrease   upon cooling below  the c phase. \color{black}
 In the field range 3.7–4 T, the susceptibility maximum $T^{\chi}_{\rm max}$ shifts to lower $T$ with increasing field. 
 When this susceptibility maximum is suppressed  at 4 T, $M(T)$ shows a remarkable enhancement of the magnetization [Fig. 3(c)].
 \color{black}

Figure 4(a) shows the $B$–$T$ phase diagram of CePdAl for $B$ $||$ $[0001]$. 
Previously, in the a phase,  a  crossover anomaly has been observed from the temperature and field dependences of the heat capacity [denoted by $T_0$ and $B_0$, respectively, in Fig. 4(a)]  \cite{Mochi_JPSJ_2017}. These anomalies are considered to  reflect the characteristic energy scale of the Kondo effect in CePdAl  \cite{Mochi_JPSJ_2017}.  When Kondo screening is suppressed in magnetic fields, the geometrical frustration cannot be lifted by the Kondo effect,   and  then   a first-order phase transition occurs abruptly at $B_{\rm ab} = $ 3.2 T.  The b phase exists only in a narrow field region, followed by another first-order phase transition at $B_{\rm bc} = $ 3.4 T. The c  phase emerges between 3.4 and 4 T and exhibits a dome-like shape as a function of $B$. 
 Figure 4(b) shows   $dM/dB$  measured at 0.58, 0.98, 1.08, 1.18, and 1.28 K. Below 1.1 K, anomalies attributable to phase transitions are observed (down arrows), whereas at 1.18 and 1.28 K a broad maximum in  $dM/dB$ appears, associated with the susceptibility maximum at 3.8 T (upper arrows). Notably, the susceptibility maximum $T_{\chi}^{\rm max}$ nearly coincides with the maximum of the c-phase dome, suggesting that frustration effects arising from the competing interactions of $J_1$ and $J_2$ \color{black}  stabilize the c phase around 4 T.
 Upon cooling,   $T_{\chi}^{\rm max} $ suddenly  disappears  near $B_{\rm cp}$ [Fig. 4(a)], suggesting that the characteristic energy scale of  dipolar frustration vanishes at  $B_{\rm cp}$.  The  occurrence of the magnetization kink and nonlinear susceptibility anomaly  at $B_{\rm p}$    suggests  that the development of nondipolar correlations is closely related to  frustration, which  remains unresolved even in the PM state above $ B_{\rm cp}$. 
  In the  higher-field  region  at $T^{*}$ and $B^{*}$, the frustration effect appears to be finally suppressed by the Zeeman splitting of the CEF Kramers doublet, where the heat capacity exhibits  Schottky-like anomalies \cite{Mochi_JPSJ_2017,Zhao,Lucas_PRL_2017}.

The possible existence  of a spin-liquid state in CePdAl has been  discussed based on its unusual magnetic phase diagram \cite{Lucas_PRL_2017, Zhang_PRB_2018};  however,   a clear physical picture has not yet been established. Lucas $et$ $al$. proposed that the spin-liquid state may emerge near the b-phase boundary \cite{Lucas_PRL_2017}, whereas Zhang $et$ $al$. discussed a  possible realization near  $B_{\rm p}$ \cite{Zhang_PRB_2018}. A quantum spin liquid is generally  regarded as a strongly correlated quantum ground state without spontaneous symmetry breaking \cite{Lee_2008, Balents}.  Our high-resolution magnetization measurements reveal   a clear thermodynamic  phase transition upon entering the b phase.  In contrast, the kink anomaly in the magnetization curve  at $ B_{\rm p}$  is a crossover  without symmetry breaking,   as demonstrated by $M(T)$ measurements   [Fig. 4(c)].

\color{black}

Let us now discuss the nonlinear-susceptibility anomaly  in CePdAl, which is manifested as a kink in the $M(B)$ curve.
 The  kink observed in the $M(B)$ curve  at $B_{\rm p}$ is reminiscent of that in Sr$_{3}$Ru$_{2}$O$_{7}$, where  
  a spin-density-wave order with  an electronic nematic correlation    emerges  near an FM quantum criticality \cite{Rost_Science_2009, Lester_NatMat_2015}. 
   \color{black}  
A similar kink in the $M(B)$ curve near a critical field has been reported in the quasi-one-dimensional compound LiCuVO$_4$, which is a strongly frustrated $S = 1/2$ system with competing AF and FM interactions \cite{Svistov_2011}. In LiCuVO$_4$, the magnetization anomaly originates from the emergence of a bond-quadrupolar (nematic) state with broken rotational symmetry \cite{Svistov_2011}. The order parameter of the nematic state is equivalent to a quadrupolar (rank-2) order
 \cite{Andreev_1984, Shannon_PRL_2006}. 
 Notably, a chiral quadrupolar order with $\bm{q} = 0$  has recently been discovered in URhSn, which also has a ZrNiAl-type structure \cite{Shimizu_PRB, Tabata_2025, Tokunaga_PRL, Kusunose_JPSJ_2024, Ishitobi}.  Importantly, the magnetization kink    anomaly in CePdAl is a crossover, as evidenced by precise $M(T)$ scans  [Fig. 4(c)]. In CePdAl, the CEF ground state is a Kramers doublet \cite{Woitschach_Neutron, Romero_CEF}, which  can be described by a pseudo-spin $1/2$. 
 Although a spin-$1/2$ system does not possess  quadrupolar degrees of freedom, a bond-quadrupolar correlation  can emerge, \color{black} driven by FM quantum fluctuations \cite{Andreev_1984, Shannon_PRL_2006}.
 Near $B_{\rm cp}$,  where the Ce(1) and Ce(3) moments are ferromagnetically polarized, the site-selective Kondo effect on the Ce(2) sites may provide a possible microscopic route to bond multipolar correlations through  anisotropic bond-dependent hopping.
Recently, the concept of augmented multipoles, including bond multipoles, has  been  intensively developed  to classify electronic degrees of freedom beyond conventional atomic multipoles \cite{Kusunose_JPhysC_2022}. 

Finally, we discuss  the Fermi-surface effects and  non-Fermi-liquid (NFL) behavior in CePdAl.
Anomalies in the Hall resistivity and thermoelectric power at $B_{\rm p}$  suggest that the 
  Fermi-surface reconstruction is associated with  the  destruction  of the Kondo  effect \cite{Zhang_PRB_2018}. 
 The Fermi-surface reconstruction  at $B_{\rm p}$ may be compatible with  the kink anomaly (i.e., change in slope)  in $M(B)$ curve. 
 However,  the expected NFL behavior   for  the simple Kondo breakdown scenario \cite{QSi_Nature_2001} is not observed  when the AF (dipole) order  is suppressed, 
  and  the magnetization $M(T)$ exhibits no  NFL behavior down to 70 mK near $B_{\rm cp}$ [Fig. 4(c)]. 
 These  findings provide important thermodynamic constraints, indicating that neither the field-induced quantum critical point associated with dipolar moments nor the Kondo breakdown scenario alone can account for the field-induced disordered state   $B_{\rm cp} < B < B_{\rm p}$ in CePdAl. 

\color{black}


In summary, we have unveiled  unusual magnetic properties in the geometrically frustrated heavy-fermion compound CePdAl using high-resolution magnetization measurements. 
We have identified three first-order phase transitions that arise from the  geometrical frustration due to the suppression of the Kondo effect in a magnetic field. Furthermore, the $M(B)$ curve at 80 mK exhibits no plateau, which is consistent with the persistence of Kondo screening  in fields up to 4 T. 
Notably, we 
 observed  a magnetization kink
  and the accompanying nonlinear susceptibility anomaly, suggesting  the presence of hidden  nondipolar  correlations   
    in the  disordered state $B_{\rm cp} < B < B_{\rm p}$.
 The anomaly at $ B_{\rm p}$   is not associated with   symmetry-breaking phase transitions.\color{black}
 We also found that the magnetization $M(T)$ exhibits  no NFL behavior when the AF  order  is suppressed near  $B_{\rm cp}$. Our  magnetization results   establish essential  thermodynamic constraints for  a unified understanding of the field-induced spin-liquid state in CePdAl. 
\color{black}


We thank T. Kanda  for valuable discussions.
The present study was supported by Grants-in-Aid KAKENHI (No. JP20K03851, JP23K03314, JP23H04870, JP23H04868) from the Ministry of Education, Culture, Sports, Science, and Technology (MEXT) of Japan. 


\begin{thebibliography}{999}
\bibitem{EndNote_inhomogeneity} We estimate the field inhomogeneity along the applied-field direction for the measured sample ($\sim$2 mm thick), which arises from the field gradient, to be only 4 mT, which is too small to account for the observed shoulder anomaly.
\end{thebibliography}


\end{document}